\documentstyle[sprocl]{article}

\def\e{{\rm e}}

\def\beq{\begin{equation}}
\def\eeq{\end{equation}}
\def\beqa{\begin{eqnarray}}
\def\eeqa{\end{eqnarray}}
\def\half{{\textstyle{1\over 2}}}
\def\pa{\partial}

\def\CE{{\cal E}}

\def\CN{{\cal N}}

\renewcommand{\tilde}{\widetilde}
\renewcommand{\det}{{\rm det}}
\renewcommand{\bar}{\overline}
\newcommand{\hf}{{1\over 2}} 

\begin{document}
\vspace*{-1cm}\hspace{10cm}{hep-th/9805062}

\title{A Proposal for Low Energy Dilaton}
\author{\vskip-8pt HoSeong La\smallskip}

\address{Department of Physics, North Dakota State Univrsity
\\ Fargo, ND 58105-5566, USA\\E-mail: hsla@golem.phys.ndsu.nodak.edu} 
\address{\vskip-8pt and\vskip-8pt}
\address{Center for Theoretical Physics, Laboratory for Nuclear Science,\\
Massachusetts Institute of Technology, Cambridge, MA 02139-4307, USA}


\maketitle\abstracts{ We propose a systematic way of introducing 
the (nongravitational) low energy dilaton and a scheme for spontaneous breaking 
of scale symmetry (SBSS) is explained.}

\vskip-.5cm
\noindent
The dilaton is a hypothetical particle which is a Goldstone boson associated
with SBSS. The earlier idea of low energy dilaton was abandoned because, in 
renormalizable QFT's, the scale symmetry suffers from the trace anomaly.
Motivated by recent interests in the gravitational dilaton, 
we decided to revisit the dilaton in the nongravity context.
If we look into the matter more carefully, the existence of anomalies
does not necessarily rule out the possibility of SSB. 
If SSB occurs in a sector different from the 
anomalous one, it is possible. A good example is the axion. 
Despite the axial anomaly, spontaneous breaking of the
Peccei-Quinn symmetry occurs so that the axion can be generated in QCD.
Furthermore, the dilaton does not transform like a quasiprimary field under
dilatations, the RG arguement does not rule out the possibility of SBSS. 

There are two key facts in our proposal. First, the dilaton is systematically
incorporated into the low energy scale invariant effective theories by 
introducing the ``dilaton geomerty'' such that the diffeomorphism (Diff) 
invariance in the dilaton geometry is equivalent to the conformal invariance of 
the Minkowski space in the presence of the dilaton. 
Second, we also need to generalize the scale symmetry by 
introducing an auxiliary scale dependent constant, hence making the scale 
symmetry ``off shell''. This is necessary because the usual scale symmetry
does not allow a potential other than the one with the runaway dilaton vacuum.
Such a constant is in some sense nothing new. For example, see the 
Wilson's approach to RG.
This talk is based on [1], in which more details can be found.

The dilaton geometry is defined by a metric of the form
\vskip-.2cm
\beq
\label{e4}
g_{\mu\nu} = \e^{2\kappa\phi}\eta_{\mu\nu},
\eeq
where $\e^{4\kappa\phi} = \sqrt{|\det g_{\mu\nu}|}$ and
$\kappa^{-1}$ is the dilaton scale. 
As pointed out in [2], if $\delta\phi$ under Diff is not as a scalar, 
but dictated by the metric, then
\vskip-.2cm
\beq
\label{e5}
\delta\e^{\kappa\phi}=\half\e^{\kappa\phi}(\pa_\mu + 4\kappa\pa_\mu\phi)v^\mu.
\eeq
Under volume-preserving diffeomorphisms (SDiff), 
$\phi$ behaves like a constant.
It is important to notice that
the consistency condition between eq.(\ref{e5}) and eq.(\ref{e4}) is
\vskip-.2cm
\beq
\label{ecnf}
\half\eta_{\mu\nu}\pa_\alpha v^\alpha =
\eta_{\mu\alpha}\pa_\nu v^\alpha + \eta_{\alpha\nu}\pa_\mu v^\alpha.
\eeq
Thus, diffeomorphisms of eq.(\ref{e4}) appear as conformal transformations of
the flat spacetime. This shows that the dilaton geometry describes the conformal
symmetry of the flat spacetime in the presence of the dilaton.
Under dilatations $v^\mu = \alpha x^\mu$, the dilaton transforms
inhomogeneously as 
\vskip-.2cm
\beq
\label{e6}
\delta\phi = \alpha\left({\textstyle{1\over\kappa}} + x^\mu\pa_\mu\phi\right),
\eeq
hence the dilaton is not a quasi-primary field. 
This distinguishes the dilaton from other fields.
Nevertheless, the dilaton is a Lorentz scalar.

Let us define a dilaton-dressed field as
$\Phi_{[d]} \equiv \e^{d\kappa\phi}\Phi$,
where $\Phi$ a scalar in the dilaton geometry. 
Then, under dilatations,
$\delta\Phi_{[d]} = \alpha (d + x^\mu\pa_\mu)\Phi_{[d]}$
so that $\Phi_{[d]}$ is a quasiprimary field.
This recovers the usual dilatations in the flat spce:
$\Phi_{[d]}(x) \to \e^{d\alpha}\Phi_{[d]}(\e^\alpha x)$ for 
$x \to \e^\alpha x$, where $d$ is the scale dimension (or the conformal weight). 
Such dressing is not needed for vector fields in four dimensions
because under dilatations
$\delta A_\mu = \alpha (1 + x^\lambda\pa_\lambda) A_\mu$.
This in particular leads to the YM term that does not couple directly to 
the dilaton in four dimensions, hence different from the Kaluza-Klein approach 
of introducing the dilaton.
In other than four dimensions vector fields still need dilaton dressing, 
leading to direct YM-dilaton couplings.
The mass dimension of a field is not necessarily the same as its scale 
dimension. For example, the dilaton has mass dimension $1$, but its scale 
dimension is not even defined.

To define the (e.g. $\CN=1$) supersymmetric dilaton geometry we need a
supersymmetric generalization of SDiff. As SDiff leaves
the volume density $\sqrt{g}$ invariant, supersymmetric SDiff
should also leave the chiral density $\CE$ invariant. 
In curved spacetime, therefore, we demand 
$\delta(\tilde{\sigma}^\mu_{\alpha\dot\alpha}\psi_\mu^\alpha)= 0$.
Using $\tilde{\sigma}^\mu_{\alpha\dot\alpha}\psi_\mu^\alpha 
= E_a^{\ \mu}\tilde{\sigma}^a_{\alpha\dot\alpha}\psi_\mu^\alpha$, we obtain
$\delta(\tilde{\sigma}^\mu_{\alpha\dot\alpha}\psi_\mu^\alpha) 
= v^\mu\pa_\mu(\tilde{\sigma}^\mu_{\alpha\dot\alpha}\psi_\mu^\alpha) 
+2\tilde{\sigma}^\mu_{\alpha\dot\alpha}\pa_\mu\zeta^\alpha = 0$.
This defines the fermionic part of the supersymmetric SDiff.
The supersymmetric dilaton geometry needs fermionic analog of eq.(\ref{e4}), 
and a good candidate is
\vspace{-.1cm}
\beq
\label{es3}
\psi_\mu^{\ \alpha} = \bar\sigma_\mu^{\dot\alpha\alpha}
\bar\psi_{\dot\alpha},
\eeq
where $\sigma$-matrices are those in flat space.
Then these supersymmetric diffeomorphisms correctly reproduce the 
superconformal transformations of flat superspace. 
Thus, defining the supersymmetric dilaton geometry, 
we identify $\psi$ as the superpartner of the dilaton, the dilatino. 
Its transformation rule under supersymmetric dilatations is 
\vskip-.2cm
\beq
\label{es5}
\delta\bar\psi_{\dot\alpha} = v^\mu\pa_\mu\bar\psi_{\dot\alpha}
+{\textstyle{1\over 4}}\bar\psi_{\dot\alpha}\pa_\mu v^\mu
-\half\sigma^\mu_{\alpha\dot\alpha}\pa_\mu
\zeta^\alpha,
\eeq
where
$2\bar\sigma_\mu^{\dot\alpha\alpha}\pa^\mu\bar\eta_{\dot\alpha}
=2\pa^\mu\pa_\mu\zeta^\alpha = \hf
\pa^\mu\pa_\mu\zeta^\alpha = 0$.
Thus the supersymmetric dilaton geometry is an effective way of 
describing the supersymmetric dilatations in the flat space.

From the supersymmetric point of view, the dilaton is always associated
with the axion. In the dilaton geometry the axion can be incorporated by 
defining
$\chi =\e^{\kappa\phi_c}$ such that $g_{\mu\nu} = \chi^*\chi\eta_{\mu\nu}$, 
where $\kappa\phi_c\equiv\kappa\phi +i {a\over f_a}$.
$\chi$ is mass-dimensionless.
Although $\chi$ is not a scalar in the dilaton geometry, 
we can demand that $\chi$ should still transform like a scale-dimension one 
field under dilatations so that the axion transformation rule can be obtained
as $\delta a = \alpha x^\mu\pa_\mu a$.
This axion can be related to the antisymmetric field $B_{\mu\nu}$ in the 
dilaton geometry.
Thus, the dilaton, axion and dilatino form a chiral supermultiplet 
$(\phi_c,\psi)$.
The R-charge of the dilaton-axion multiplet can be easily determined from 
this and the superconformal symmetry such that its R-charge is zero.
This enables us to define dressing for fermions.
Fermions should be dressed over each Weyl components such that
$\Psi_{[d]} \equiv {\psi_{1[d]\alpha}\choose \bar\psi_{2[d]}^{\dot\alpha}}
= \left({\chi^d\atop 0} {0\atop {\chi^*}^d}\right)
{\psi_{1\alpha}\choose \bar\psi_{2}^{\dot\alpha}}
\equiv \left({\chi^d\atop 0} {0\atop {\chi^*}^d}\right)\Psi$.

We demand a superconformally invariant action in the presence of the
dilaton-axion multiplet to be covariant in the supersymmetric dilaton geometry. 
Thus, such an effective action can be read off from the supergravity action 
using the metric of the supersymmetric dilaton geometry. 
The bosonic case can be analogously defined too. 
Introducing a dynamical scale which transforms under dilatations as
$\delta M = \alpha M$,
hence making the scale transformation ``off-shell'',
we can in fact write down a dilaton effective potential which is free from
a trace anomaly: 
A generic form of the pure dilaton part of a scalar potential which 
exhibits SBSS is
\vskip-.3cm
\beq
\label{e35}
V_{{\rm dil, eff}} = \kappa^2\Lambda\e^{4\kappa\phi}(\phi - a_1)(\phi -a_2),
\eeq
where $\Lambda$ is a mass dimension four constant and
$\kappa a_i = \half\ln{\kappa^2\Lambda\over M^2} +c_i, \ i=1,2,$
and $c_i$'s are constants independent of $M$. The specific forms of $a_i$
are dictated to make the potential scale invariant so that there is no
trace anomaly from this potential.
This scalar potential can be derived from a superpotential, for example,
with one scalar field, $z$, other than the dilaton,
\vskip-.5cm
\beq
\label{e37}
W_{{\rm eff}}(\phi_c, z) = 
{\textstyle{1\over 2\sqrt{3}}}\sqrt{\Lambda}(a_1-a_2)\e^{3\kappa\phi_c} + 
\sqrt{\Lambda}\kappa z\e^{2\kappa\phi_c}\left(\phi_c -\half(a_1+a_2)\right).
\eeq
The first term is an R-symmetry breaking term that vanishes as $(a_1-a_2)\to 0$.
$V_{{\rm dil,eff}}$ has a scale symmetry breaking vacuum at 
\vskip-.3cm
\beq
\langle\phi\rangle = v \equiv \half\left(a_1+a_2 -{\textstyle{1\over 2\kappa}} 
+\sqrt{(a_1-a_2)^2 + {\textstyle{1\over 4\kappa^2}}}\right)
\eeq
with $V_{{\rm dil,eff}}(\langle\phi\rangle)\leq 0$. 
The equality is only for $a_1 = a_2$.
The dilaton mass in the nonsymmetric vacuum reads
$m_\phi^2 = 2\kappa^2\Lambda\e^{4\kappa v}\left(1 + 2\kappa(2v -a_1 -a_2)
\right)$.
As $v\to -\infty$, $m_\phi\to 0$, confirming the dilaton is massless in the
scale invariant asymptotic vacuum.
To be more precise, the vacuum must be a vacuum for the total scalar potential 
$V_{{\rm eff}}$, not just the pure dilaton part $V_{{\rm dil}}$. 
The vacuum structure of $V_{{\rm eff}}$ is fairly complicated even with just 
one chiral multiplet. There are
three different types of vacua with SBSS: unbroken supersymmetry, 
spontaneously broken supersymmetry and softly broken supersymmetry. 
If the scale symmetry is broken in the bosonic vacuum,  then the Poincar\'e 
supersymmetry must be broken at the same time. If the scale symmetry is broken 
in the fermionic vacuum but the bosonic vacuum remains invariant, then the 
Poincar\'e supersymmetry can be preserved as long as the R-symmetry breaking is 
specifically related to the scale symmetry breaking

If $a_1 = a_2$, there is no axion contribution to the scalar potential
other than the axion mass term in my examples. Otherwise, however, 
the scalar potential explicitly contains nontrivial axion potential in 
some cases.
The axion effective potential roughly takes the form of
\vskip-.6cm
\def\af{{\textstyle{a\over f_a}}}
\beqa
V_{{\rm eff}}(a)&=&f(\phi, z, \bar{z}) a^2 + g(\phi, z, \bar{z})(a_1-a_2) 
\nonumber\\
&&\times\left[{\rm Re}z\!\left(h(\phi)\cos\af
+\af\sin\af\right)\!-{\rm Im} z\left(\af\cos\af
-h(\phi)\sin\af\right)\right]\nonumber
\eeqa
for smooth functions $f$, $g$ and $h$. In particular, the potential is not 
periodic. The first term is due to the term necessary for scale symmetry 
breaking, hence the dilaton contribution breaks the periodicity of the axion
potential. The nonperiodicity in the second term is because the corresponding 
term in the superpotential does not have R-symmetry. 
Nevertheless, it still yields an  extremum at $a = 0$. The periodic
axion potential can be obtained in the limit $h(\phi)\sim\phi\to -\infty$
for $\phi\gg a$, which is the scale invariant limit, and
$z$ gets a real vacuum expectation value.
All the vacua I obtain determine $\langle a\rangle =0$.

If we incorporate such a structure into the gravitational case such that
the low energy dilaton is related the gravitational dilaton, 
the Diff symmetry of curved spacetime may appear as spontaneously
broken down to SDiff symmetry. Although this still contains the 
Poincar\'e symmetry, hence, nothing against current experimental 
observations, there is much prejudice against such an idea. 
We hope nature herself will clarify this in future.

\smallskip
\noindent
{\bf Acknowledgements:} I thank R. Arnowitt, 
D.Z. Freedman, M. Grisaru and G. Mack for helpful correspondence.
I am also much grateful to B. Zwiebach at MIT and R. Hammond at NDSU 
for their hospitality while this work was done.
This work is supported in part by DOE and NASA.

\smallskip
\noindent
{\bf References}\par
1. H.S. La,  hep-th/9608148, hep-th/9704196.\par
2. H.S. La, hep-th/9510147.




\end{document}